\documentclass{article}
\usepackage{authblk}
\usepackage{graphicx}
\usepackage{amsmath,amssymb}
\usepackage{csquotes}

\title{Single-cycle scalable terahertz pulse source in refleciton geometry}

\author[1,*]{Gy\"orgy T\'oth}
\author[1]{L\'aszl\'o P\'alfalvi}
\author[1]{Zolt\'an Tibai}
\author[1]{Levente Tokodi}
\author[2,3]{J\'ozsef A. F\"ul\"op}
\author[1]{Zsuzsanna M\'arton}
\author[1,2]{G\'abor Alm\'asi}
\author[1,2,3]{J\'anos Hebling}

\affil[1]{Institute of Physics, University of P\'ecs, 7624 P\'ecs, Hungary}
\affil[2]{Szent\'agothai J\'anos Research Centre, University of P\'ecs, 7624 P\'ecs, Hungary}
\affil[3]{MTA-PTE High-Field Terahertz Research Group, University of P\'ecs, 7624 P\'ecs, Hungary}

\affil[*]{Corresponding author: tothgy@fizika.ttk.pte.hu}

\begin{document}
\maketitle

\textbf{A tilted-pulse-front pumped terahertz pulse source is proposed for the generation of extremely high field single-cycle terahertz pulses. The very simple and compact source consists of a single crystal slab having a blazed reflection grating grooved in its back surface. Its further important advantages are the energy scalability and the symmetric THz beam profile. Generation of 50 MV/cm focused field with 5.6 mJ terahertz pulse energy is predicted for a 5 cm diameter LiNbO$_3$ crystal, if the pump pulse is of 450 mJ energy, 1030 nm central wavelength and 1 ps pulse duration. Such sources can basically promote the realization of THz driven electron and proton accelerators and open the way for a new generation concept of terahertz pulses having extreme high field.}

\section{Introduction}

Acceleration of electrons \cite{Nanni:2015,Zhang:2018} and protons \cite{Palfalvi:2014,Sharma:2016} are promising applications of THz pulses with extremely high (>100 kV/cm) electric field. Optical rectification of ultrashort laser pulses in nonlinear materials (NMs) can be appropriate for efficient generation of such THz pulses. The highest so far THz pulse energy (0.9 mJ) was achieved by using the organic DSTMS as NM \cite{Vicario:2014}. However, the spectrum obtained from organic materials is typically centered in the 2 to 10 THz, while lower frequencies are more suitable for particle acceleration. The frequency range below 2 THz can be better accessed with lithium niobate (LiNbO$_3$, abbreviated as LN in the followings), utilizing the tilted-pulse-front (TPF) technique for non-collinear phase matching \cite{Hebling:2002}. Presently, THz pulse energy in the mJ level is available with TPF excited THz sources using LN \cite{Fulop:2014}.

In the TPF excitation geometry, the velocity matching reads as 
\begin{equation}
    v_{p,gr}\cos\left(\gamma\right) = v_{THz,ph},
    \label{eq:velocity_matching}
\end{equation}
where $v_{p,gr}$ is the group velocity of the pump pulse, $v_{THz,ph}$ is the phase velocity of the THz pulse, and $\gamma$ is the pulse-front-tilt angle, which is $\sim$63$^{\circ}$ for LN.

LN based TPF sources have been demonstrated to be very effective and, because of this, are widely applied in THz pump-probe experiments supporting mJ level THz pulses. However, the large necessary tilt-angle and the correspondingly large angular dispersion of the pump beam mean strong limitations, namely: (i) Imaging errors in the presence of angular dispersion result in a lengthened pulse duration at the edges of the pump spot \cite{Fulop:2010,Kunitski:2013}. (ii) In order to enable perpendicular incoupling of the pump and perpendicular outcoupling of the THz beam, a prism-shaped nonlinear crystal has to be used with a wedge angle equal to the tilt angle $\gamma$. This results in THz pulses with different temporal shapes at the edges of the THz beam because of the different generation lengths. Such a bad quality, strongly asymmetric THz beam drastically hinders many applications, especially particle acceleration. (iii) Due to the large group-delay-dispersion (GDD) \cite{Hebling:1996} associated to the angular dispersion, an ultrashort pump pulse evolves very fast inside the LN crystal, and the average pulse duration becomes much longer than the Fourier transfrom-limited (FL) pump pulse duration.

Problems (i) and (ii) can be eliminated, in principle, using contact-grating set-ups \cite{Palfalvi:2008}. An efficient contact-grating set-up was realized using ZnTe semiconductor as nonlinear material \cite{Fulop:2016}. Contact grating setups were designed and realized for LN, too \cite{Ollmann:2012,Tsubouchi:2014}. However, because of technical difficulties induced by the large needed tilt angle, efficient generation of THz pulses has failed yet. 

In order to mitigate problem (iii) one should use semiconductor crystal, which requires lower pulse-front-tilt, or longer FL pump pulse duration can be applied \cite{Fulop:2010}. As another solution, the effect of the angular dispersion can be avoided by using stair-step echelon structure. In this case a stair-step shaped, segmented tilted-pulse-front is formed instead of a continuous one. The first suggestion and demonstration of such a source type was published in Ref. \cite{Ofori:2016}. Efficient generation of THz pulses by 70 fs long pump pulses were demonstrated by using a reflective echelon \cite{Ofori:2016}. Disadvantageously, however, this echelon setup also requires both imaging and a prism-shaped NM with the same large $\gamma$ wedge angle as the conventional setup.
 
Recently we suggested a so called “nonlinear echelon slab” (NLES) THz source, which is a NM with an echelon-like profile created on its entrance surface.  This configuration has the advantage that a slab-like NM can be used, and that it also reduces the imaging errors. High THz conversion efficiency with symmetric, homogeneous THz beam can be achieved  if appropriate pre-tilt is applied in the pump beam \cite{Palfalvi:2017,Nugraha:2019}. Although the slab-like structure allows significant energy scalability, imaging errors still remain a limiting factor. Simulations for a modified NLES setup predicts good THz conversion efficiency in an imaging free configuration as well \cite{Toth:2019}. Disadvantageously, it is necessary that the NLES has a small wedge angle.  

In the present paper we propose a very simple, compact, energy-scalable THz pulse source that avoids limiting factors (i) and (ii). A blazed reflective grating structure is created on the backside of a nonlinear slab. High diffraction efficiency can be achieved with this structure in high diffraction order. Neither a pre-tilt of the pump beam, nor imaging optics have to be used. High THz energies with excellent THz beam properties are predicted with this simple and compact setup.

\section{The RNLS THz source}

The presently proposed THz source consists of only the pump source and a reflective nonlinear slab (RNLS). Here we emphasize, that this compactness is a very important advantage of the setup. Fig. 1 depicts the RNLS itself. It is made of a nonlinear crystal (LN is supposed). Its front (top) surface is plane, and its back (bottom) surface is structured in order to form a blazed reflective grating. The front surface and normal of the grating on the back surface are parallel to each other. According to the geometry of the garting structure, we distinguish two different implementations (symmetric and asymmetric). We focus on the symmetric one in the present paper. The asymmetric version might also have relevance, but we do not discuss it here.

\begin{figure}
\centering
\includegraphics[width=8cm]{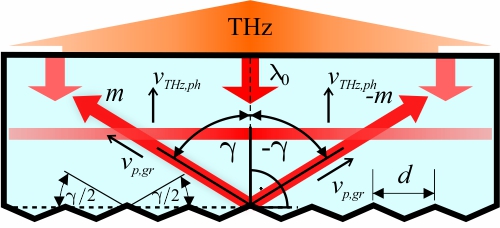}
\caption{A schematic figure of the reflective nonlinear slab (RNLS) THz source. The red color concerns the pump.}
\label{fig:setup}
\end{figure}

The pump beam incident perpendicular to the top surface of the slab (see Fig. \ref{fig:setup}) does not generate practically any THz radiation when it propagates inside the NM towards the back surface, since velocity matching condition is not fulfilled. According to the 
\begin{equation}
    \sin(\alpha)+\sin(\beta) = m\frac{\lambda}{dn_{p,ph}}
    \label{eq:grat_eq}
\end{equation}
grating equation for perpendicular incidence ($\alpha=0$) the $\pm$ orders are arranged symmetrically to the normal of the grating (similarly as in the contact grating scheme examined in \cite{Bakunov}). If the grating structure is symmetric, the beams of the +m and –m orders have the same intensity. The necessary velocity matching can be achieved in a tilted-pulse-front geometry. The diffraction angles belonging to the chosen $\pm m$ orders have to be equal to the velocity matching angle as $\beta_m=\gamma$ and $\beta_{-m}=-\gamma$. This can be achieved by adequate choice of the $m/d$ ratio. In both the +m$^{th}$ and -m$^{th}$ order beam a tilted front are formad, which \textbf{arte} on each other and are parallel to the front surface (the red line in Fig. \ref{fig:setup}.), and its velocity component in the direction of the normal of the grating is $v_{p,gr}\cos\left(\gamma\right) = v_{THz,ph}$, satisfying the velocity matching condition. The generated THz pulse propagates perpendicularly to the tilted-pulse-front i.e. also perpendicularly to the front surface of the RNLS. Hence, the THz beam exits from the slab without deflection.    

In order to obtain large diffraction efficiency, the grating has to be blazed with blaze angle equal to $\gamma/2$. It was shown by COMSOL simulations, that high diffraction efficiency can be achieved in high diffracion orders. The diffraction efficiency together with the necessary grating constant belonging to the given $\pm m$ diffraction order-pair can be seen in Fig. \ref{fig:racs}a for 800 and 1030 nm pump wavelengths. We have examined the diffraction efficiency for two cases, namely for gold reflection coating on the back side of the RNLS (LN-gold) and in the absence of any coating (LN-air), since both solutions might have practical relevance. LN-air boundary was supposed in the THz generation simulations. It is seen in Fig. \ref{fig:racs}. that the efficiency approaches the $\sim$80\%  saturation value for $m > 20$, therefore in the calculations we supposed $m=20$. The necessary grating constant (for 800 and 1030 nm, LN-air) is also shown in Fig. \ref{fig:racs}a. In order to have information on the diffraction efficiency for broad pump pulses Fig. \ref{fig:racs}b shows the wavelength dependence of the diffraction efficiency for the $m=20$ diffraction order as an example.  

Since there is spatial overlap between the incident pump and the outcoming THz beams, they have to be separated. It can be done by applying a (e.g. $\mbox{InO}_2$) dichroic beam splitter, or by slightly tilting out the slab from the orientation perpendicular to the propagation direction of the incoming pump. 

Similarly to Ref. \cite{Nugraha:2019}, we suggest mechanical micromachining by diamond tool for the manufacturing of the periodic structure.

\section{Numerical model}

A simple model was developed in order to obtain quantitative information on the optical-to-THz conversion efficiency and on the THz pulse shape and spectra for the reflective slab source introduced in Section 2. The model described in Ref. \cite{Vodopyanov:2006} for THz generation by optical rectification was adapted to the proposed new setup.  The noncollinear excitation geometry, as well as the variation of the pump pulselength with the propagation due to the angular dispersion were taken into consideration in the model.

Let us suppose a Gaussian initial temporal shape of the optical pump pulse: $E(t)=\mathrm{Re}\left[E_0\exp\left(-2\ln\left(2\right)t^2/\tau_0^2\right)\exp\left(i\omega_0t\right)\right]$, where $E_0$ is the peak electric field-strength an $\tau_0$ is the FL pump pulse length (FWHM). Taking into account the nonlinear polarization in the same way as in Ref. \cite{Vodopyanov:2006}, the differential equation for the $E_{THz}\left(\Omega,z'\right)$ Fourier component of the THz field reads as:

\begin{eqnarray}
   &&\frac{\partial}{\partial z'}E_{THz}\left(\Omega,z'\right) = \\ \nonumber
    &&-i\frac{\Omega d_{eff}E_0^2\tau_0}{4\sqrt{\pi\ln(2)}c n_{THz,ph}(\omega)}\exp\left[-\frac{\left(\tau_0^2+\Delta\tau(z')^2\right)\Omega^2}{16\ln(2)}\right]\times \\ \nonumber
   &&\exp\left[i\frac{\Omega}{c}\left(n_{THz,ph}(\Omega)-\frac{n_{p,gr}}{\cos(\gamma)}\right)z'\right]-\frac{\alpha(\Omega)}{2}E_{THz}(\Omega,z'),
   \label{eq:differential_equation}
\end{eqnarray}
where $z'$ is the THz propagation coordinate, $c$ is the speed of light in vacuum, $\Omega$ is the angular frequency of the THz radiation, $n_{THz,ph} = c/v_{THz,ph}$ is the THz phase index of refraction, $\alpha_{THz}(\Omega)$ is the THz (intensity) absorption coefficient, $d_{eff}$ is the effective nonlinear coeffcient ($d_{eff}= 168\,\, \mathrm{pm}/\mathrm{V}$ for LN), and $n_{p,gr}$ is the group refractive index of the pump. The relation between the $z'$ THz propagation coordinate and the $z_p$ pump coordinate is
\begin{equation}
    z' = z_p\cos(\gamma).
    \label{eq:z_vesszo}
\end{equation}

One way to increase the effective interaction length (defined in \cite{Fulop:2010}) is to compensate for the effects of material and angular dispersion (see the first exponential term of Eq. (\ref{eq:grat_eq}), by pre-chirping the pulses \cite{Fulop:2010, Bakunov}. In this way, one can set the pump-pulse length to be minimal (i.e., equal to the FL value) not at the entrance, but inside the crystal by tuning the prechirp. Instead of following the complex optimization procedure given in \cite{Bakunov}, we set the FL pulse length value simply at the crystal center (at $z'=L/2$, where $L$ is the thickness of the slab) \cite{Fulop:2010}. 

In Eq. (\ref{eq:grat_eq}) $\Delta\tau(z')$ is the delay between the short and long wavelength components at the half of the spectral intensity maximum of the pump pulse at $z'$. Starting from the dispersion parameter, for the case of simultaneous presence of angular and material dispersion \cite{Hebling:1996}, one can obtain

\begin{equation}
    \Delta\tau(z') = \frac{\lambda_0}{c}\frac{z'-L/2}{\cos(\gamma)}\left(\frac{n_{p,gr}^2}{n_{p,ph}}\frac{\tan^2(\gamma)}{\lambda_0^2}-\frac{\mathrm{d}^2n_{p,ph}}{\mathrm{d}\lambda^2}\right)\Delta\lambda,
    \label{eq:tau}
\end{equation}
where $\Delta\lambda$ is the spectral bandwidth (FWHM). As can be seen $\Delta\tau(L/2)=0$ resulting the FL pump pulselength at the middle of the crystal as was aimed.

The second exponential factor in Eq. (\ref{eq:grat_eq}) takes into account any remaining velocity mismatch coming from the effect of the dispersion in the THz range. Terahertz absorption, which is taken into account in the very last term of Eq. (\ref{eq:grat_eq}), is related only to the lattice. Absorption by free-carriers, generated by multiphoton absorption is neglected. Any nonlinear effects (except the optical rectification), such as pump depletion, and self-phase modulation were also neglected.

\begin{figure}
\centering
\includegraphics[width=\linewidth]{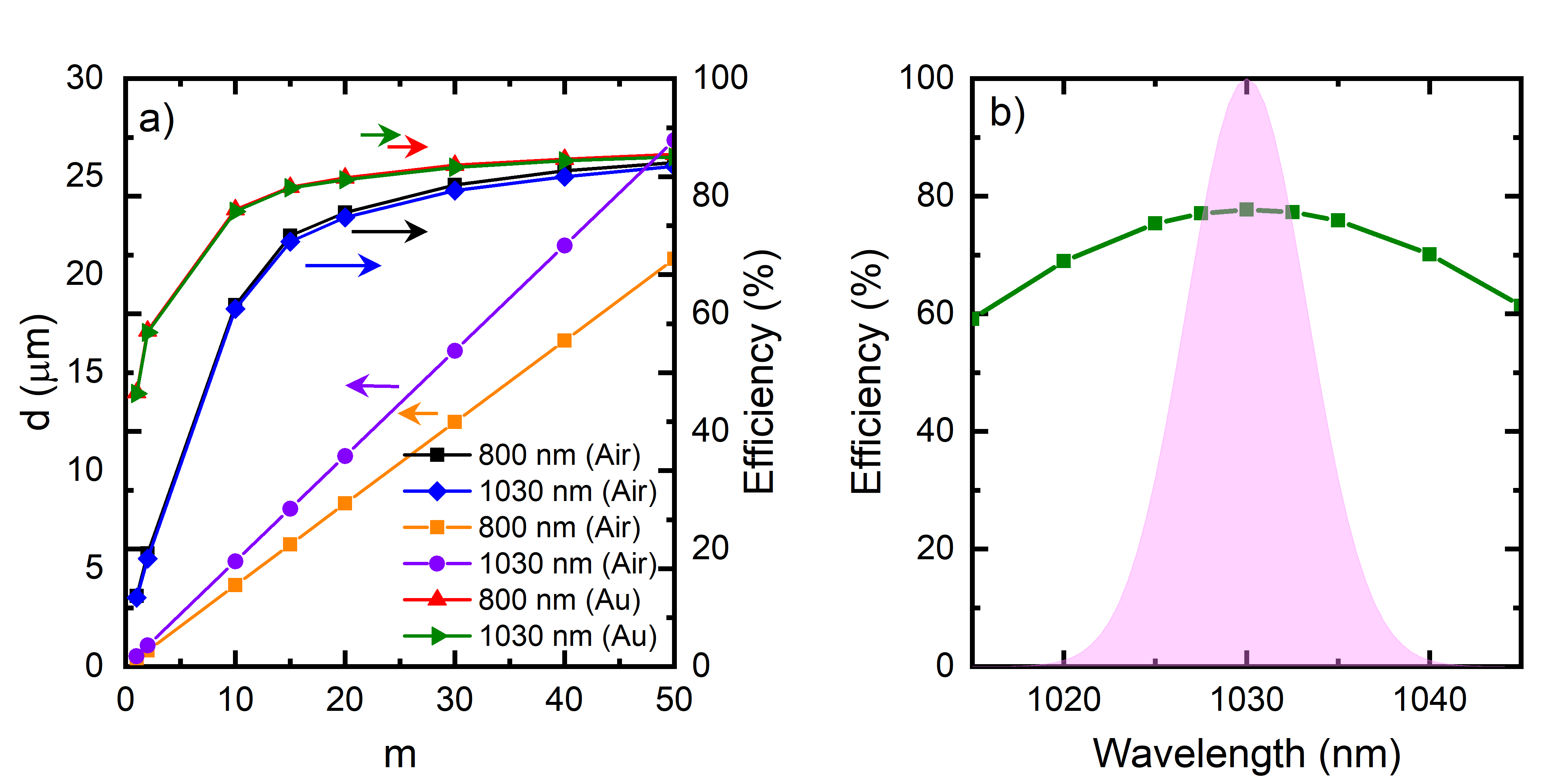}
\caption{Diffraction efficiency and grating period versus the diffraction order for 800 and 1030 nm wavelengths (a). The wavelength dependence of the diffraction efficiency for the $m=20$ diffraction order and the spectrum of a 200 fs FL pump pulse for reference.}
\label{fig:racs}
\end{figure}

The THz fluence at the output can be determined from the solution of Eq. (\ref{eq:grat_eq}) as
\begin{equation}
    Fluence_{THz}=\frac{\epsilon_0 c}{2}2\pi\int_0^\infty\left|E_{THz}(\Omega,L)\frac{2n_{THz,ph}}{n_{THz,ph}+1}\right|^2\mathrm{d}\Omega.
\end{equation}
Fresnel reflections, which cause a significant loss, are also accounted for by this formula. The pump fluence is:
\begin{equation}
    Fluence_p = \sqrt{\frac{\pi}{2}}\frac{\epsilon_0n_{p,ph}}{2}E_0^2\frac{\tau_0}{\sqrt{2\ln(2)}}
\end{equation}
and the optical-to-THz conversion efficiency is:
\begin{equation}
    \eta = \frac{Fluence_{THz}}{Fluence_p}\eta_D,
\end{equation}
where $\eta_D$ is the diffraction efficiency of the reflection grating (see Fig. \ref{fig:racs}).

A 0.68 mol\% doped stoichiometric LN was supposed as NM, and the temperature was supposed to be 300 K in the simulations. Two widely used laser types were assumed as pumping sources, which differ not only in their central wavelength but also in the typical pulse lengths. One of the considered pump sources worked at 800 nm (Ti:Saphire lasers), delivering pulses in the 50 - 1500 fs range, while the other worked at 1030 nm (Yb-doped lasers), delivering pulses in the 200 - 1500 fs range. The damage
threshold intensity of crystals is approximately proportional to $1/\sqrt{\tau_0}$ \cite{Stuart:1995}. Therefore, the pump intensity was chosen according to 
\begin{equation}
I_0=I_0^* \sqrt{\frac{100\mathrm{fs}}{\tau_0}},
\label{eq:intensity(tau)}
\end{equation}
where $I_0^* = 100 \frac{\mathrm{GW}}{\mathrm{cm}^2}$ is the peak pump intensity at 100 fs. In this way, the used pump intensity is about four times lower than the damage threshold on the whole investigated pump pulselength range. By choosing relatively low pump intensity many nonlinear effetcts are irrelevant, and the B-integral (concerning the pump propagation through the crystal before reaching the grating) remains below 0.5. 

\section{Results}

\begin{figure}
\centering
\includegraphics[width=\linewidth]{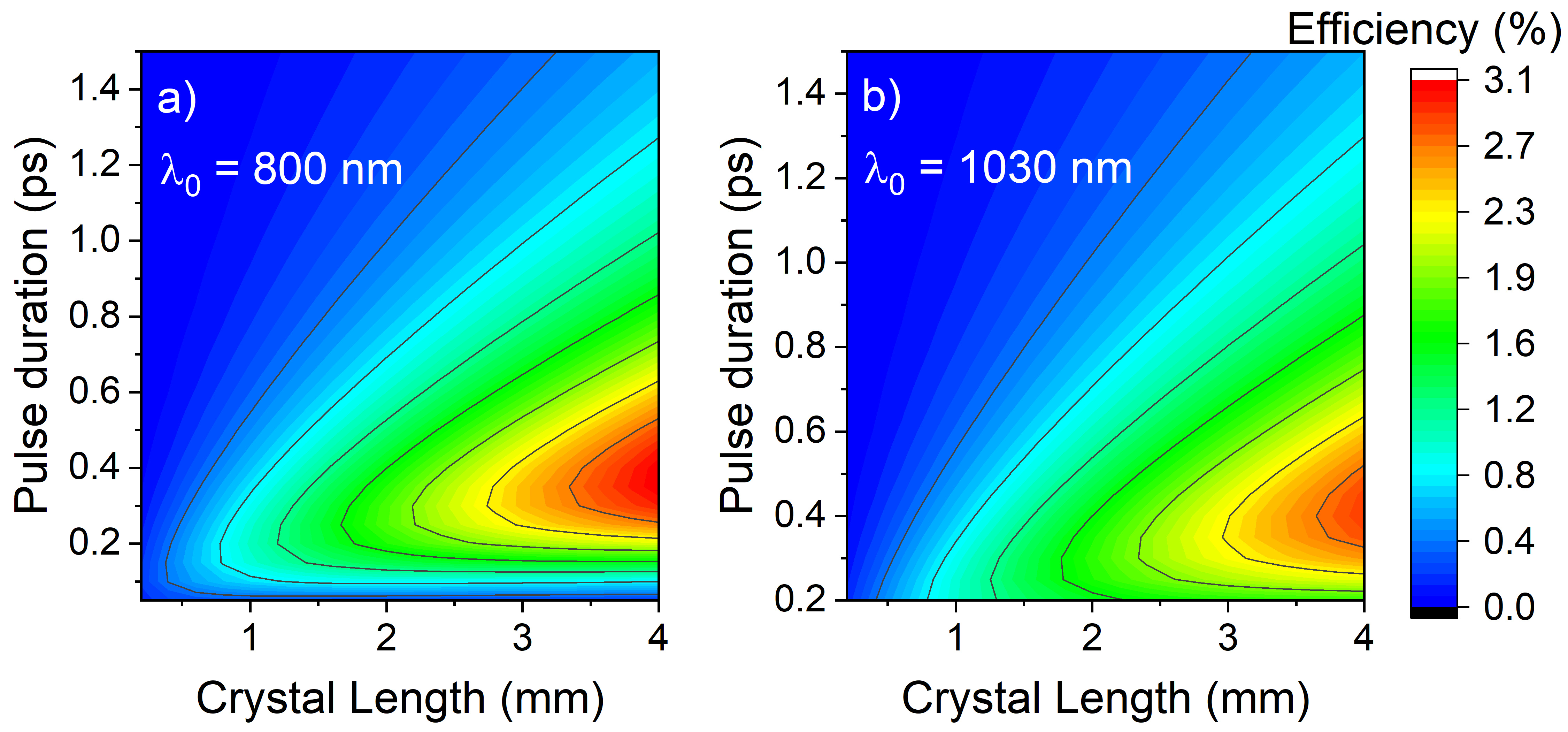}
\caption{Optical-to THz conversion efficiency as a function of the pump pulse length and the crystal length for 800 (a) and 1030 nm (b) pump wavelengths.}
\label{fig:effic}
\end{figure}

Fig. \ref{fig:effic} shows the optical-to THz conversion efficiency predicted for the RNLS source as a funtion of the pump pulse length and the crystal length for 800 and 1030 nm pump wavelengths. Maximal efficiency can be achieved at 4 mm crystal length and $\sim$400 fs pump pulse length in the examined region. 

Higher efficiency can be reached at 800 nm, than at 1030 nm. The reason for this is that for shorter wavelengths the variation of the pump pulse length with the pump propagation is slower, resulting in longer interaction length \cite{Fulop:2010}.


As it is seen in Fig. \ref{fig:effic}, the optimal pump pulse length to reach maximal efficiency increases with the crystal length. For example the optimal FL pump pulse length is 200 fs for 1 mm, but 400 fs for 4 mm crystal length at 1030 nm pump wavelength. The pump peak intensity is larger for shorter FL pump pulselength according to Eq. \ref{eq:intensity(tau)}, but it should not be forgotten that the pump pulse length  increases (Eq. \ref{eq:tau}), and hence the pump intensity decreases rapidly during the propagation due to large angular dispersion. This results in the shortening of the interaction length, so the output does not scale linearly with the peak pump intensity. Consequently, if the FL pump pulse of the available pump source is longer it is practical to use longer crystals. 

\begin{figure}
\centering
\includegraphics[width=\linewidth]{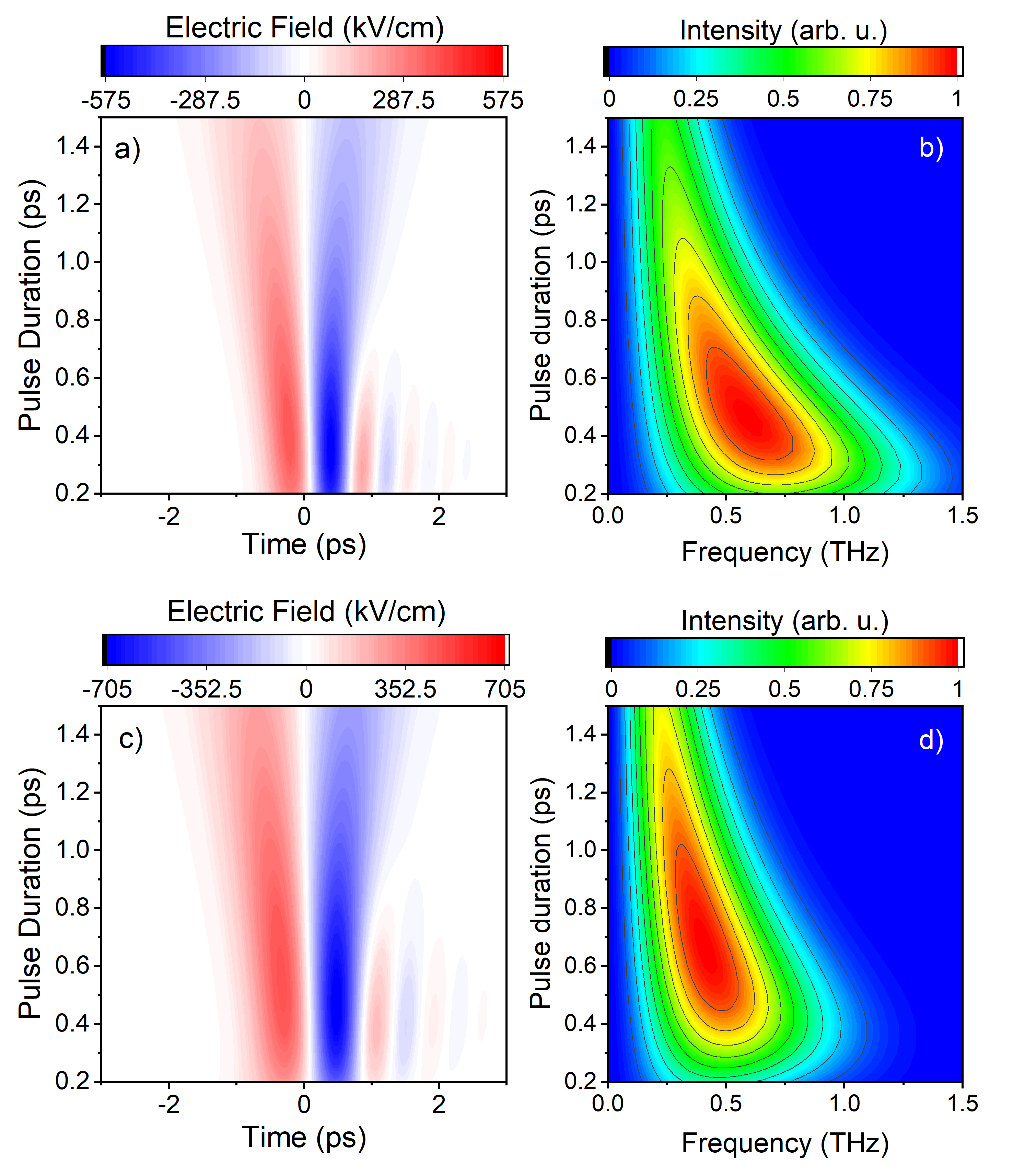}
\caption{THz pulseforms (a,c) and the corresponding spectra (b,d) for 2 (a,b) and 4 mm (c,d) crystal lengths.}
\label{fig:contourshape}
\end{figure}

It is worth to determine the temporal characteristics of the electric field strength, which are important for several applications, especially particle acceleration. Fig. \ref{fig:contourshape} shows the THz pulseforms (a,c) and the corresponding spectra (b,d) for 2 (a,b) and 4 mm (c,d) crystal lengths at 1030 nm pump wavelength. The peak electric field strength is 575 kV/cm for a 2 mm crystal, while it is 700 kV/cm for 4 mm length. The spectral peak shifts towards lower frequencies as the crystal length gets longer, since the absorption coefficient of the LN increases with the THz frequency.  


As it is seen in Fig. \ref{fig:contourshape}, the main THz cycle is followed by several oscillations for short FL pump pulses. The reason for this is the material dispersion of the LN in the THz range. This explanation is supported by Fig. \ref{fig:contourshape} (b,d). The THz spectrum becomes broader for short pump pulses. Due to the large spectral band   width, the material dispersion has significant influence on the temporal pulse shape, a chirp is introduced.  THz absorption results in narrow spectrum in long crystals (Fig. \ref{fig:contourshape}d). However, the effect of the dispersion remains significant because of the long propagation distance. Accordingly, it follows that the appearance of the oscillations can not be significantly influenced by the crystal length.   

\begin{figure}
\centering
\includegraphics[width=\linewidth]{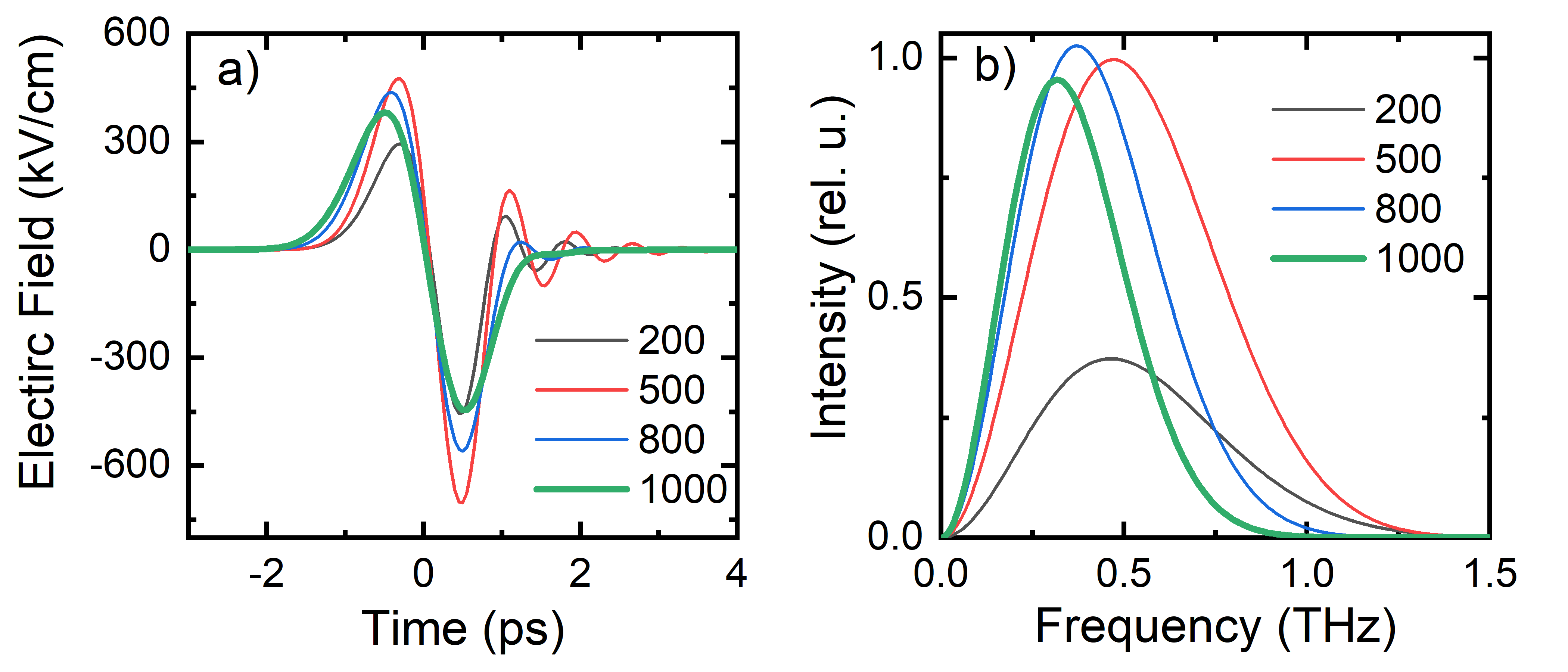}
\caption{THz pulseshapes (a) and the corresponding spectra (b) for several FL pump pulselengts supposing 4 mm crystal length.}
\label{fig:shape}
\end{figure}

As it is evident from Fig. \ref{fig:contourshape}, THz pulses become perfectly single-cycle for FL pump pulse lengths longer than 1 ps.  The central frequency of the THz pulse decreases with increasing pump pulse duration. For the sake of clarity, Fig. \ref{fig:shape} shows a few selected pulse shapes with their corresponding spectra. 



\begin{figure}
\centering
\includegraphics[width=\linewidth]{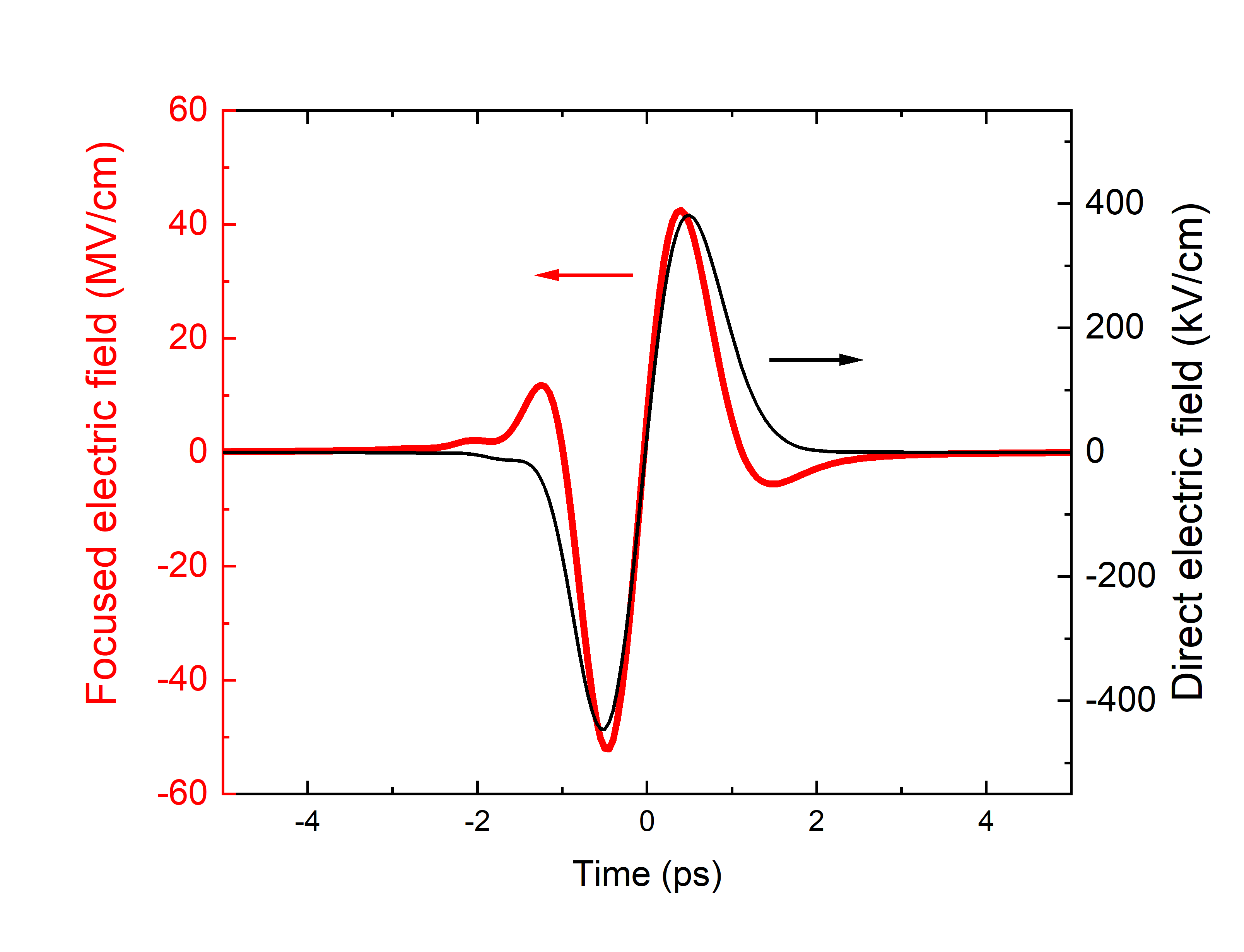}
\caption{Terahertz pulse shapes directly after exiting the RNLS (black) and after focusing (red) supposing 4 mm crystal length, 1 ps FL pump pulse length, 5 cm beam diameter, 2.5 cm focal length (NA$\approx 1$). The pulse energies of the pump and the THz are 450 and 5.6 mJ, respectively. }
\label{fig:focused}
\end{figure}

Beside its simplicity, the other main impotrant advantage of the RNLS source is that there is no principal limitation for its excitation surface. Consequently, there are no principle limitations for the scalability of the THz pulse energy. Extreme large electric field strength can be achieved by focusing THz pulses excited by RNLS having large surface. For the demonstration of this, we supposed pump pulses with 1 ps FL pulse length, 450 mJ energy,  5 cm beam diameter, 1030 nm wavelength and crystal length of 4 mm. According to Fig \ref{fig:effic}, the optical-to-THz generation efficiency is 1.25, $\%$ i.e. the THz pulse energy is 5.6 mJ in this case.  50 MV/cm peak electric field strength can be obtained by focusing the generated 5 cm diameter THz beam with an off axis parabolic mirror of 2.5 cm focal length.  Fig. \ref{fig:focused} shows the THz pulse shapes directly behind the RNLS and in the focus. Although the pulse shape becomes somewhat distorted by  focusing \cite{Hunsche:1999,Feng:1998}, since the foci of the spectral components of the broad THz spectra are spatially separated, the focused field remains nearly single-cycled. This property makes this THz pulse extremely suitable for particle acceleration. Our results suggest that over 1 ps pump pulse length even longer than 4 mm crystal lengths may be applicable. 


\section{Discussion}

The proposed RNLS TPF THz pulse source overcomes limitations (i) and (ii), namely the imaging errors, and the necessity of the use of the prism shaped NM. Although the angular dispersion (iii) is not reduced, for pump pulses longer than 200 fs the setup predicts good optical-to-THz generation efficiencies even for pump intensities well below the damage threshold, and for relatively short crystals.

\begin{figure}
\centering
\includegraphics[width=60mm]{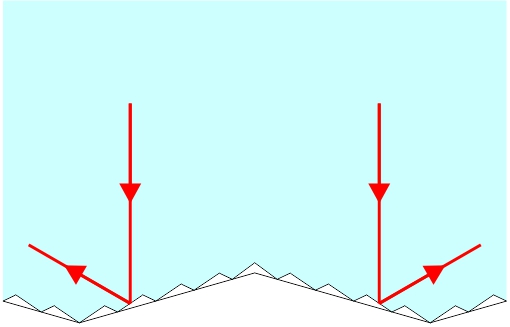}
\caption{The double structure RNLS consisting of a rough and a fine structure.}
\label{fig:vegyes struktura}
\end{figure}

As it was pointed out in the Introduction, the setup suggested by Ofori et al.  \cite{Ofori:2016} is free of angular dispersions, but is limited by the factors (i) and (ii).  In order to reduce the effect of angular dispersion in the above discussed RNLS arrangement, we propose to use it in high ($m\gtrsim 200$) diffraction order. This means that the $d$ grating constant also has to be increased significantly, so practically the RNLS works as a ribbed reflector. Such a setup can be applied, if the time delay between the neighbouring zones is larger than the pump pulse length. The weakness of this solution is the destructive THz interference coming from the segmented pump pulse front. This can be overcome by a double structure, which is a combination of a rough and a fine structure as shown in Fig. \ref{fig:vegyes struktura}. Similarly to the hybrid property of the NLES \cite{Palfalvi:2017}  the necessary average pulse front tilt is also created by both diffraction and reflection in this structure, but here the diffractive and reflective elements are integrated. The theoretical modeling and quantitative analysis of this altered RNLS structure are not the subject of the present paper, since deeper considerations are needed.

All the previously discussed results belong to simulations at 300 K.  We found that the efficiencies are nearly doubled at 100 K, due to the lower THz absorption of the LN. The optimal crystal length for obtaining single cycle pulses is shorter for 100 K than for 300 K, while the optimal pump pulse length is larger for 100 K than for 300 K. This behavior of the parameters can be explained by the fact that the THz spectrum shifts towards higher frequencies due to the lower THz absorption at lower temperatures.   

It is important to note, that simulated efficiency values exceeding 2\% are not reliable, because of several limiting effects due to the presence of intense THz field as discussed in \cite{Ravi:2014}. Such effects might be taken into account in a more complex model in the future. The goal of the present paper was to show the advantages of the proposed RNLS setup in the frame of a simple model. These advantages are the scalability of the THz pulse energy and the extremely large available peak electric field strength together with spatially uniform THz beam properties. 

LN was used as NM in the above discussed simulations. However, the advantages of the RNLS setup can also be exploited using semiconductor as NM.

\section{Conclusion}
We introduced an imaging free, slab shaped THz pulse source, which allows the upscaling of the THz pulse energy without principal limitations.The temporal characteristics of the electric field strength is uniform along the whole cross section of the THz beam independently of the beam spot size. Perfect single cycle pulses with efficiency exceeding 1\% can be generated by FL pump pulses of ~1 ps length. By the focusing of these pulses  several 10 - 100 MV/cm peak electric field strength can be achieved, depending on the beam spot size. These properties open the way for several novel applications of high energy THz pulses. For exampe, such pulses can play important role in the field of THz-based particle accelaration \cite{Nanni:2015,Palfalvi:2014,Sharma:2016}. 



\bibliographystyle{unsrt}
\bibliography{main}



\end{document}